\documentclass[aip,pop,amsmath,amssymb,reprint]{revtex4-2}

\usepackage{graphicx}
\usepackage{dcolumn}
\usepackage{bm}
\usepackage{amsmath}
\usepackage{amssymb}
\usepackage{subcaption}
\usepackage{hyperref}
\usepackage[dvipsnames]{xcolor}

\begin{document}

\title{Differentiable Programming for Plasma Physics: From Diagnostics to Discovery and Design}

\author{A. S. Joglekar}
\email{archis@ergodic.io}
\affiliation{Ergodic LLC, Seattle, WA 98103}
\affiliation{Pasteur Labs, Brooklyn, NY 11205}
\affiliation{Laboratory for Laser Energetics, University of Rochester, Rochester, NY 14623}
\affiliation{Department of Nuclear Eng. and Rad. Sciences, University of Michigan, Ann Arbor, MI 48109}

\author{A. G. R. Thomas}
\affiliation{Department of Nuclear Eng. and Rad. Sciences, University of Michigan, Ann Arbor, MI 48109}
\affiliation{Gerard Mourou Center for Ultrafast Optical Sciences, University of Michigan, Ann Arbor, MI 48109}

\author{A. L. Milder}
\affiliation{Laboratory for Laser Energetics, University of Rochester, Rochester, NY 14623}

\author{K. G. Miller}
\affiliation{Laboratory for Laser Energetics, University of Rochester, Rochester, NY 14623}

\author{J. P. Palastro}
\affiliation{Laboratory for Laser Energetics, University of Rochester, Rochester, NY 14623}

\author{D. H. Froula}
\affiliation{Laboratory for Laser Energetics, University of Rochester, Rochester, NY 14623}

\date{\today}

\begin{abstract}
Differentiable programming, enabled by automatic differentiation (AD), provides a robust framework for gradient-based optimization in computational plasma physics. While optimization is often only used towards design, we demonstrate that it can also be used for discovery and bridging the gap towards multi-scale modeling. We discuss four applications: (1) discovering novel nonlinear plasma phenomena, including a previously unknown superadditive wavepacket interaction regime, by optimizing differentiable kinetic simulations; (2) learning hidden variables that capture spatiotemporally non-local kinetic effects in fluid simulations, enabling hydrodynamic models to reproduce large Knudsen number physics typically requiring kinetic solvers; (3) accelerating Thomson scattering analysis by over $140\times$ while enabling extraction of velocity distribution functions with $\mathcal{O}(10^3)$ parameters; and (4) inverse design of spatiotemporal laser pulses that achieve target far-field behavior where full space-time coupling improves performance by $15\times$ over spatial or temporal optimization alone. These examples illustrate that differentiable programming not only accelerates existing design and inference workflows but enables qualitatively new capabilities, from algorithmic physics discovery to high-dimensional inference and design previously considered intractable.
\end{abstract}

\maketitle

\section{Introduction}

Data-driven modeling has become increasingly prominent in plasma physics \cite{hatfield_data-driven_2021}, driven by the high computational cost of the multiscale nature of plasma dynamics, and the growing complexity of experimental diagnostics. Data-driven approaches are now routinely employed for surrogate modeling \cite{djordjevic_modeling_2021}, reduced-order modeling \cite{kaptanoglu_physics-constrained_2021, barbour_machine-learning_2025}, equation discovery \cite{alves_data-driven_2022}, and experimental design \cite{gaffney_making_2019, spears_predicting_2025}. At the same time, there is broad recognition that purely data-driven approaches face fundamental challenges in extrapolation, interpretability, and consistency with known physical constraints.

These problems are not unique to plasma physics. In the context of fluid dynamics and turbulence modeling, Brenner \emph{et al.}~\cite{brenner_perspective_2019} emphasize the need for approaches that balance data-driven flexibility with mechanistic structure, arguing that learned models should complement rather than replace established physical theory. Similar perspectives have emerged across scientific computing, where surrogate models and closure relations are widely used but must be embedded within physically meaningful frameworks to remain predictive outside their training regimes  \cite{kochkov_machine_2021,shankar_differentiable_2023}.

Recent work in differentiable modeling and scientific machine learning has proposed unifying perspectives on this balance\cite{sapienza_differentiable_2024}. Shen \emph{et al.}~\cite{shen_differentiable_2023} frame scientific models as compositions of analytical operators and learned components, emphasizing differentiability as the key enabling property that allows these elements to be combined, optimized, and interrogated systematically. In parallel, Rackauckas \emph{et al.}~\cite{rackauckas_universal_2021} introduce the concept of universal differential equations (UDEs), in which unknown or unresolved physics is represented by universal function approximators embedded directly within differential equation solvers. In this view, classical simulations, learned closures, and fully data-driven models arise as limiting cases within a single computational framework.


\begin{table*}
\caption{Taxonomy of data-driven approaches in computational plasma physics, organized by what the learned component replaces and where physical knowledge enters.}
\label{tab:taxonomy}
\begin{ruledtabular}
\begin{tabular}{lllll}
\textbf{Approach} & \textbf{What is learned} & \textbf{Physics via} & \textbf{Strengths} & \textbf{Limitations} \\
\hline
Surrogate model & Full input--output map & Training data & Fast inference & No guarantees beyond training \\
Neural operator & Solution operator & Architecture & Resolution-invariant & Physics not encoded \\
PINN & Solution representation & Loss function & Mesh-free; flexible & Optimization can be fragile \\
Diff.\ simulation & Selected components & Discrete operators & Interpretable; extrapolates & Requires diff.\ solver \\
Equation discovery & Equation structure & Candidate library & Human-readable laws & Needs well-chosen basis \\
\end{tabular}
\end{ruledtabular}
\end{table*}

Building on these ideas, it is useful to distinguish data-driven approaches in plasma physics according to \emph{what the learned component replaces} in the computational pipeline and \emph{where physical knowledge enters}. Table~\ref{tab:taxonomy} summarizes five broad categories.

Surrogate models learn the entire input--output mapping from data, with physical structure enforced only implicitly through the training distribution. They are widely used in plasma physics to accelerate parameter scans and enable real-time inference, but they carry no structural guarantee outside the training regime. Neural operators~\cite{kovachki_neural_2023} share this fully learned character but embed mathematical structure in the form of spectral convolutions or resolution invariance that attempts to mimic the architecture of physics-based solvers. This architectural inductive bias improves generalization relative to generic surrogates, yet the governing equations themselves still do not appear in the model.

Physics-informed neural networks (PINNs)\cite{raissi_physics-informed_2019} also learn a full solution representation but incorporate physical knowledge by penalizing the residual of the governing equations in the loss function. Constraints are thus enforced at the level of optimization rather than through a discrete forward model, which provides flexibility at the cost of optimization difficulty in stiff or multiscale systems.

Differentiable simulation takes a fundamentally different approach. The discretized governing equations and numerical solvers remain the core computational object. Automatic differentiation is applied directly to the discrete simulation pipeline, enabling exact derivatives with respect to model parameters, initial conditions, or latent variables. Learning, when present, is \emph{surgically localized} so that unknown components such as closure relations or transport coefficients may be represented by neural networks, while the surrounding physics-based structure is preserved. Traditional simulations correspond to the limiting case in which no components are learned.

Finally, equation discovery methods such as sparse regression\cite{rudy_data-driven_2017,alves_data-driven_2022,kaptanoglu_permanent-magnet_2022, kaptanoglu_sparse_2023} and symbolic regression operate at a different level entirely: rather than learning solutions or parameters, they learn the \emph{form} of governing equations from data. The output is not a prediction but a human-readable mathematical expression. These methods are complementary to differentiable simulation because once a candidate equation is discovered, it can be embedded within a differentiable solver for validation and refinement. We return to this connection in Sec.~\ref{sec:conclusion}.

The key distinction between differentiable simulation and the other learned approaches is where the physics \emph{lives}. In surrogates, neural operators, and PINNs, the neural network \emph{is} the solver because it produces the solution directly, with physics entering only through data, architecture, or loss penalties. In differentiable simulation, the physics-based solver produces the solution, and the neural network merely generates parameters or closures \emph{for} that solver. This difference has practical consequences for interpretability, conservation, and extrapolation that are illustrated throughout the applications in Secs.~\ref{sec:kinetic_discovery}--\ref{sec:design}.

\section{Automatic Differentiation: Concepts and Tradeoffs}

Automatic differentiation (AD) is a set of techniques for computing exact derivatives of numerical algorithms by systematically applying the chain rule to elementary operations \cite{griewank_evaluating_2008}. Unlike symbolic differentiation, AD operates on the computational graph of a program rather than closed-form expressions, and unlike finite differences, it introduces no truncation error beyond that of the underlying floating-point arithmetic. A comprehensive treatment can be found in the survey by Baydin \textit{et al.}~\cite{baydin2018}. This section reviews the key concepts and their implications for plasma physics applications.

\subsection{Computational graphs and accumulation modes}

In plasma physics, the function to be differentiated is often defined implicitly through a numerical solver and can be expressed as
\begin{equation}
\mathbf{y} = \mathbf{f}(\mathbf{x}),
\end{equation}
where $\mathbf{x}$ may represent physical parameters, initial conditions, or control inputs, and $\mathbf{y}$ denotes simulation outputs or diagnostic observables. Automatic differentiation constructs derivatives of $\mathbf{f}$ by augmenting the numerical evaluation rather than manipulating governing equations symbolically. This distinction is critical for plasma simulations, where the mapping from inputs to outputs may involve millions of timesteps, nonlinear solvers, and complex boundary conditions that would be intractable to differentiate analytically.

Any numerical algorithm decomposes into a sequence of elementary operations forming a computational graph. AD associates additional quantities with each intermediate variable and propagates them according to the chain rule. In \emph{forward-mode} AD, each variable carries a tangent representing its directional derivative with respect to a chosen input. In \emph{reverse-mode} AD, each variable carries an adjoint representing the sensitivity of a chosen output with respect to that variable. Both modes differentiate the same primal computation but traverse the computational graph in opposite directions.


\subsection{Computational scaling and memory requirements}

\begin{table*}[t]
\caption{Comparison of derivative computation methods. $N$ denotes the number of input parameters and $M$ the number of outputs.}
\label{tab:derivative_methods}
\begin{ruledtabular}
\begin{tabular}{llllll}
Method & Accuracy & Cost scaling & Memory & Implementation effort & Best suited for \\
\hline
Finite differences & Approximate & $\mathcal{O}(N)$ & Low & Very low & Prototyping \\
Analytic adjoint & Exact (continuous) & $\mathcal{O}(1)$ & Moderate--high & Very high & Fixed solvers \\
Forward-mode AD & Exact (discrete) & $\mathcal{O}(N)$ & Low & Low & Low-dim.\ sensitivities \\
Reverse-mode AD & Exact (discrete) & $\mathcal{O}(M)$ & High (checkpointable) & Low--moderate & Large inverse problems \\
\end{tabular}
\end{ruledtabular}
\end{table*}

The distinction between forward and reverse mode has direct computational consequences. For a function $\mathbf{f}: \mathbb{R}^N \to \mathbb{R}^M$, forward-mode AD computes Jacobian--vector products efficiently but requires $\mathcal{O}(N)$ passes to assemble a full gradient. Reverse-mode AD computes vector--Jacobian products, evaluating the gradient of a scalar objective at cost proportional to $M$ but largely independent of $N$.

For the inverse problems central to this article, or to machine learning in general, the objective is typically a scalar loss function ($M = 1$) while the number of parameters $N$ may range from tens to millions. Reverse-mode AD is therefore the method of choice because it enables the calculation of the full gradient in a single backward pass regardless of $N$. This scaling advantage is summarized in Table~\ref{tab:derivative_methods} and underlies all applications discussed in subsequent sections.

The primary cost of reverse-mode AD is memory. The backward pass requires access to intermediate states from the forward evaluation. For time-dependent plasma simulations, this may include the full temporal history of fields and distribution functions. For multi-dimensional fluid simulations and kinetic simulations, this can often exceed available GPU memory.

\emph{Checkpointing} addresses this by trading computation for storage~\cite{griewank_algorithm_2000}. Rather than storing all intermediate states, the forward computation is divided into segments with only boundary states retained; intermediate values are recomputed as needed during the backward pass. Optimal strategies reduce memory from $\mathcal{O}(T)$ to $\mathcal{O}(\sqrt{T})$ for $T$ timesteps at the cost of at most one additional forward evaluation. All kinetic simulations discussed in Sections ~\ref{sec:kinetic_discovery} and \ref{sec:closures} employ checkpointing.

\subsection{Relation to adjoint methods in plasma physics}

Using adjoint methods for sensitivity analysis and optimization is not new for plasma physicists. Stellarator design codes use adjoints to compute gradients of confinement metrics with respect to coil geometries~\cite{paul_adjoint_2020,paul_gradient-based_2021} where the adjoint provides sensitivity of a scalar objective to high-dimensional inputs at cost comparable to a single forward simulation.

Reverse-mode AD can be understood as an automatically generated discrete adjoint. A manually derived adjoint of a discretized solver, if implemented correctly, matches reverse-mode AD applied to that solver. The advantage of AD is automation because the adjoint is generated by the framework rather than derived by hand.

This automation becomes critical as codes grow in complexity. Traditional adjoint methods require substantial effort. The analysis-first ``optimize-then-discretize'' approach demands careful treatment of boundary conditions, while ``discretize-then-optimize'' requires differentiating every operation including time integrators, nonlinear solvers, and limiters. AD handles these structures automatically, provided the code is implemented in a compatible framework such as JAX~\cite{jax2018github} or PyTorch~\cite{Paszke}.

\subsection{From parameter optimization to function learning}
\label{sec:workflow_evolution}

\begin{figure}[t]
\centering
\includegraphics[width=0.5\textwidth]{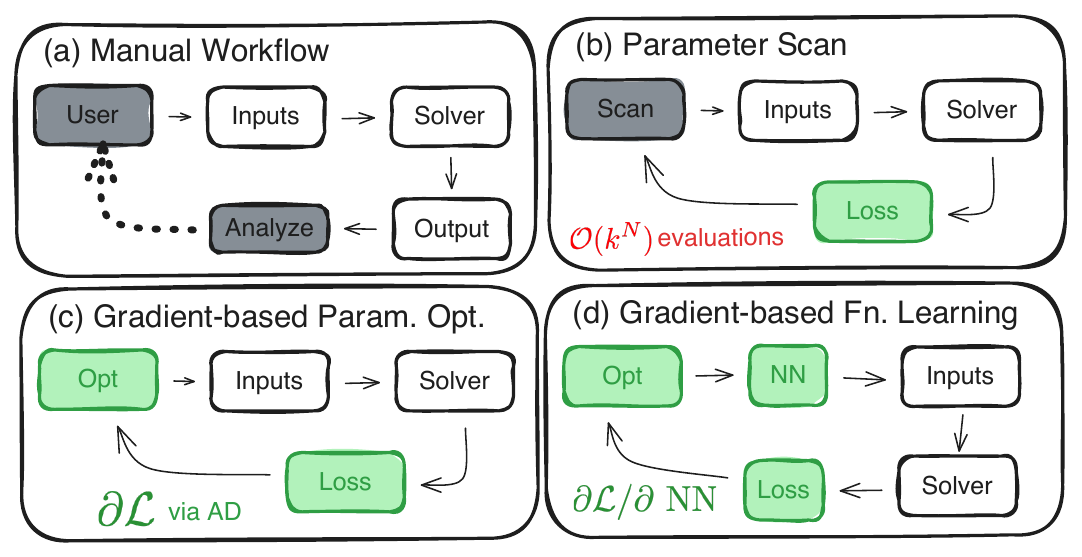}
\caption{Computational workflows enabled by differentiable programming. (a) Manual iteration. (b) Parameter scan with cost scaling $\mathcal{O}(k^N)$. (c) Gradient-based optimization using reverse-mode AD. (d) Function learning through a neural network embedded in a differentiable solver. The four applications in this article span stages (c) and (d).}
\label{fig:workflow}
\end{figure}

Beyond computational efficiency, automatic differentiation enables a qualitative shift in how physics problems can be formulated. Figure~\ref{fig:workflow} illustrates this evolution through four stages.

In the \emph{manual workflow} (Fig.~\ref{fig:workflow}a), a researcher specifies parameters, runs a simulation, analyzes outputs, and iterates by hand. This approach does not scale. For example, understanding a four-dimensional parameter space at modest resolution requires thousands of simulations and substantial human effort.

\emph{Parameter scanning} (Fig.~\ref{fig:workflow}b) automates iteration by introducing a scalar cost function $\mathcal{L}$ that quantifies agreement with a target. However, brute-force scanning scales as $k^N$ for $k$ samples along each of $N$ dimensions, becoming intractable for $N > 3$.

\emph{Gradient-based optimization} (Fig.~\ref{fig:workflow}c) exploits derivative information to navigate efficiently toward optima. With reverse-mode AD providing exact gradients at $\mathcal{O}(1)$ cost, this approach scales to hundreds of parameters. The diagnostic analysis in Section~\ref{sec:thomson} operates in this regime, fitting plasma conditions and instrumental parameters to measured spectra.

\emph{Function learning} (Fig.~\ref{fig:workflow}d) represents the most powerful configuration. Here, the parameters $\mathbf{p}$ fed to the solver are not optimized directly but are generated by a neural network. This can be expressed mathematically as
\begin{equation}
\mathbf{p} = \mathcal{G}(q; \boldsymbol{\theta}),
\end{equation}
where $q$ denotes physical inputs (e.g., wavenumber, excitation time) and $\boldsymbol{\theta}$ are the network weights. The loss function depends on the solver output, $\mathcal{L} = \mathcal{L}(\mathcal{V}(\mathbf{p}))$, and gradients propagate through the entire composition so that the gradient is given by 
\begin{equation}
\frac{\partial \mathcal{L}}{\partial \boldsymbol{\theta}} = 
\frac{\partial \mathcal{L}}{\partial \mathcal{V}}
\frac{\partial \mathcal{V}}{\partial \mathcal{G}}
\frac{\partial \mathcal{G}}{\partial \boldsymbol{\theta}}.
\end{equation}

This formulation learns \emph{functions} rather than discrete optimal values. The neural network encodes continuous relationships between physical inputs and optimal control parameters, relationships that may be difficult or impossible to derive analytically. Crucially, the network does not replace the physics. Instead, it generates parameters for a first-principles solver whose structure provides physical constraints and interpretability. The kinetic discovery application (Section~\ref{sec:kinetic_discovery}) and closure learning application (Section~\ref{sec:closures}) both operate in this regime.

\subsection{Implications for plasma physics}

The $\mathcal{O}(1)$ scaling of reverse-mode AD has several immediate consequences for plasma physics:

\textbf{High-dimensional optimization becomes tractable.} Traditional parameter studies require $k^N$ simulations; gradient-based optimization typically converges in $\mathcal{O}(N)$ to $\mathcal{O}(N^2)$ iterations regardless of resolution. This enables optimization over hundreds or thousands of parameters, sufficient to represent continuous functions (Section~\ref{sec:kinetic_discovery}), hidden dynamical variables (Section~\ref{sec:closures}), or full velocity distribution functions (Section~\ref{sec:thomson}).

\textbf{Inverse problems can be posed within simulations.} Because AD differentiates through the solver itself, the objective can depend on any computed quantity. Questions such as ``What initial perturbation maximizes an observable?'' or ``What collision operator reproduces observed relaxation?'' become concrete optimization problems with computable gradients.

\textbf{Uncertainty quantification improves.} Exact gradients enable efficient Hessian computation, which characterizes curvature near optima. Under appropriate assumptions, the inverse Hessian provides parameter covariance estimates. Section~\ref{sec:thomson} applies this to Thomson-scattering analysis.

\textbf{Learning replaces assumption.} Empirical closures, fitted transport coefficients, and ad hoc prescriptions can be replaced with flexible, learnable representations whose form is determined by data rather than assumed a priori. Section~\ref{sec:closures} demonstrates this for kinetic closures in fluid models.

In the sections that follow, we demonstrate these capabilities through three applications spanning theoretical discovery, computational modeling, and experimental analysis. Finally, Section VI applies the same framework to inverse design, optimizing near-field laser structure to achieve desired far-field behavior including uniform ionization in nonlinear media.

\section{Discovering Novel Kinetic Physics Through Differentiable Simulation}
\label{sec:kinetic_discovery}

Automatic differentiation enables a reformulation of a core activity in theoretical plasma physics---finding novel dynamical behavior---as an inverse problem. Rather than prescribing driver parameters and interpreting simulation outputs post hoc, one directly optimizes parameters, or entire functions of parameters, using physically motivated objective functionals. This section summarizes such an approach applied to nonlinear kinetic plasma physics, based on Joglekar and Thomas~\cite{joglekar_unsupervised_2022}.

\subsection{From parameter scans to function learning}

Traditional computational physics workflows proceed as follows: a researcher specifies initial conditions and forcing parameters, runs a simulation, analyzes the output, and iterates manually. When searching for optimal parameters, this becomes a brute-force scan that scales as $k^N$ for $k$ values along each of $N$ dimensions.

Differentiable programming enables a qualitative shift in this workflow. By implementing the simulation in an AD-compatible framework, the scalar objective becomes a differentiable function of the input parameters, and gradient-based optimization replaces exhaustive search. More powerfully, the parameters themselves can be generated by a neural network, allowing the optimization to learn \emph{functions} rather than discrete optimal values. This yields continuous, interpretable relationships between physical inputs and optimal control parameters.

\subsection{Governing equations and differentiable formulation}

We consider the one-dimensional Vlasov--Poisson--Fokker--Planck (VPFP) system,
\begin{equation}
\frac{\partial f}{\partial t}
+ v \frac{\partial f}{\partial x}
- E \frac{\partial f}{\partial v}
=
\left( \frac{\partial f}{\partial t} \right)_{\mathrm{coll}},
\label{eq:vpfp}
\end{equation}
with the self-consistent electric field determined from Gauss's law,
\begin{equation}
\frac{\partial E}{\partial x}
=
1 - \int f \, dv.
\label{eq:gauss}
\end{equation}
Here $f(x,v,t)$ is the electron distribution function, and the collision operator represents electron-electron collisions with $\nu_{ee}/\omega_p = 5 \times 10^{-5}$, corresponding to inertial fusion conditions. 

The solver is implemented in JAX, an AD-compatible framework, rendering the full time evolution differentiable with respect to all input parameters. Gradient checkpointing is employed at every timestep to manage memory costs, reducing storage requirements from $\mathcal{O}(T)$ to $\mathcal{O}(\sqrt{T})$ at the cost of recomputing intermediate states during the backward pass.

\subsection{Inverse problem formulation}

Let $\mathcal{V}$ denote the differentiable VPFP solver mapping initial conditions and control parameters $\mathbf{p}$ to a final state, and let $\mathcal{C}$ denote a scalar objective functional. The inverse problem is
\begin{equation}
S = \mathcal{C}(f_f) = \mathcal{C}(\mathcal{V}(f_0, \mathbf{p})).
\label{eq:inverse_problem}
\end{equation}
Automatic differentiation yields exact gradients $\partial S / \partial \mathbf{p}$, enabling efficient optimization.

To move beyond optimizing individual parameters, the control parameters are generated by a neural network:
\begin{equation}
\mathbf{p} = \mathcal{G}(q; \boldsymbol{\theta}),
\end{equation}
where $q$ denotes physical inputs and $\boldsymbol{\theta}$ are the network weights. The chain rule gives
\begin{equation}
\frac{\partial S}{\partial \boldsymbol{\theta}}
=
\frac{\partial \mathcal{C}}{\partial \mathcal{V}}
\frac{\partial \mathcal{V}}{\partial \mathcal{G}}
\frac{\partial \mathcal{G}}{\partial \boldsymbol{\theta}},
\end{equation}
allowing the network to learn continuous functional relationships from physics-based objectives alone, without labeled training data. The key conceptual step is that the neural network does not replace the physics but rather learns relationships \emph{through} the differentiable physics model.

\subsection{Application: learning optimal wavepacket excitation strategies}

The framework is applied to the interaction of nonlinear electron plasma wavepackets. Large-amplitude Langmuir waves occur in inertial confinement fusion through stimulated Raman scattering (SRS). When such waves trap electrons, the trapped population suppresses Landau damping. However, because resonant electrons travel at the phase velocity $v_{\mathrm{ph}} = \omega/k$, which exceeds the group velocity $v_g = \partial \omega / \partial k$ by a factor of 3--5, these electrons transit through the wavepacket from back to front. Their departure from the rear allows Landau damping to resume there, eroding (``etching'') the wavepacket from back to front and limiting the duration of high-field structures~\cite{fahlen_propagation_2009}.

The behavior of the transiting electrons remained unexamined. Specifically, whether it was possible for a second wavepacket to be excited such that it interacts constructively with the electrons streaming from a first wavepacket? We pose this as an optimization problem. A first wavepacket is excited with fixed parameters $(x_0, \omega_0, t_0, k_0)$. Given the wavenumber $k_0$ and a desired excitation time $t_1$ for a second wavepacket, we seek to learn the optimal spatial location $x_1(t_1, k_0)$ and frequency shift $\Delta\omega_1(t_1, k_0)$.

These quantities are parameterized by a neural network (2 layers, 8 nodes wide, with Leaky-ReLU activations) and optimized using a physics-motivated loss function given by
\begin{equation}
\mathcal{L} = -U_{\mathrm{es}} - \Delta S_K,
\label{eq:loss}
\end{equation}
where
\begin{equation}
U_{\mathrm{es}} = \int_{t_i}^{t_f} \!\! \int E^2 \, dx \, dt
\end{equation}
is the time-integrated electrostatic energy, and
\begin{equation}
\Delta S_K = \int_{t_i}^{t_f} \!\! \int \!\! \int
\left( f \ln f - f_M \ln f_M \right) dv \, dx \, dt
\end{equation}
measures the cumulative deviation from a local Maxwellian $f_M$. Minimizing this loss simultaneously maximizes electrostatic energy (favoring persistent waves) and non-Maxwellianness (favoring strongly nonlinear kinetic configurations). The key insight is that this objective encodes no explicit knowledge of resonant electron dynamics, it simply rewards configurations far from equilibrium with strong fields.

Training is performed over a parameter space with $k_0 \in [0.26, 0.32]$ and $t_1 \in [400, 800]\,\omega_p^{-1}$, comprising 35 input samples. Each simulation runs on a grid of $N_x = 6656$, $N_v = 160$ points for $1100\,\omega_p^{-1}$. Training converges in approximately 60 epochs (2100 total simulations, 150 GPU-hours on an NVIDIA T4).

\subsection{Results: discovery of superadditive wavepacket interactions}

\begin{figure}[t]
    \begin{subfigure}{0.5\textwidth}
        \includegraphics[width=\textwidth]{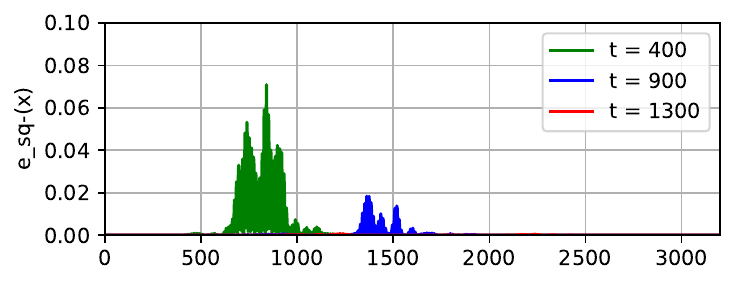}
    \end{subfigure} 
    \\
    \begin{subfigure}{0.5\textwidth}
        \includegraphics[width=\textwidth]{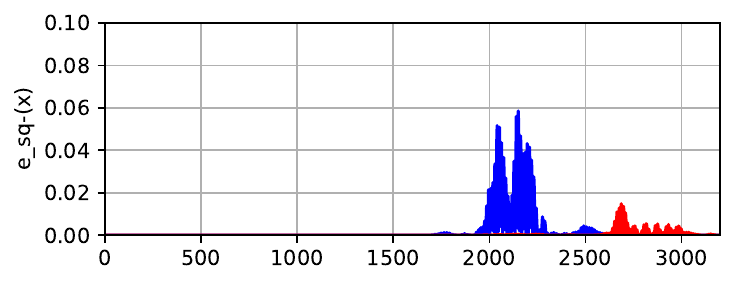}
    \end{subfigure}
    \\
    \begin{subfigure}{0.5\textwidth}
        \includegraphics[width=\textwidth]{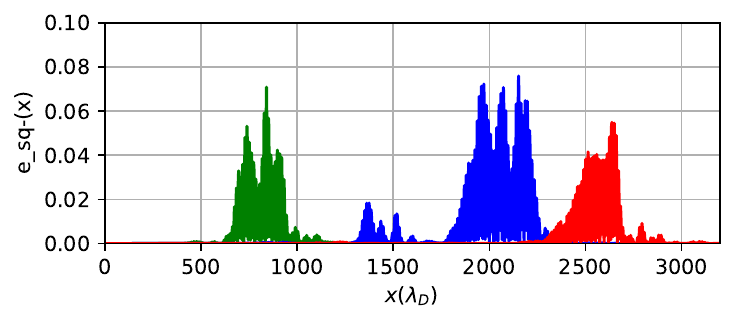}
    \end{subfigure}
\caption{Superadditive wavepacket interactions discovered through optimization. Electric field amplitude $E^2(x)$ at t = 400, 900, and 1300 $\omega_p^{-1}$ for (a) first wavepacket only, (b) second wavepacket only, and (c) both with optimized excitation parameters. Late in time, the isolated second wavepacket has damped away while the combined system retains substantial energy—the whole exceeds the sum of its parts. From Joglekar and Thomas~\cite{joglekar_unsupervised_2022}.}
\label{fig:superadditive}
\end{figure}

The optimization discovers a regime in which the combined two-wavepacket system sustains electrostatic energy far longer than either wavepacket in isolation, a \emph{superadditive} effect where the whole exceeds the sum of its parts, $f(x+y) > f(x) + f(y)$, as shown in Fig.~\ref{fig:superadditive}.

Phase-space analysis reveals the underlying mechanism. Electrons detrapped from the first wavepacket stream forward at $v_{\mathrm{ph}}$ and arrive at the rear of the second wavepacket. There, they flatten the velocity distribution near the phase velocity, suppressing the re-emergence of Landau damping that would otherwise etch the second wavepacket. The second wavepacket thus propagates freely while continuously receiving streaming electrons from the first.

Crucially, the learned functions recover physically interpretable scalings that were not imposed \emph{a priori}:
\begin{itemize}
\item The optimal excitation location follows $x_1 \approx v_{\mathrm{ph}} \, t_1$, consistent with resonant electron transport at the phase velocity.
\item The learned frequency shift $\Delta\omega_1 \sim 0.01$--$0.06$ increases with wavenumber, matching the expected scaling from trapped-particle theory.
\item The optimization identifies a ``resonance surface'' in $(x, t)$ space represented by a locus of points where second wavepackets can be excited to produce long-lived structures.
\end{itemize}

These relationships emerged purely from gradient-based optimization of the physics-based objective, demonstrating that differentiable simulations can discover mechanisms, not merely optimal parameters.

\subsection{Practical considerations}


The primary computational challenge is memory. Differentiating through $N_t$ timesteps naively requires storing all $N_t \times N_x \times N_v$ distribution function snapshots. Gradient checkpointing, implemented via JAX's built-in rematerialization, reduces this to storing only checkpoint states at the cost of one additional forward pass. For longer simulations or higher-dimensional problems, adjoint-based approaches that solve a backward-in-time ODE for the gradient may offer further memory savings~\cite{chen_neural_2019}.

\subsection{Discussion}

This example demonstrates that differentiable programming can transform theoretical plasma physics from forward exploration to inverse problem solving. The learned functions provide interpretable physical relationships, not black-box surrogates, because the neural network acts only as a function generator while all dynamics are governed by the VPFP equations. The approach is data-efficient (35 training samples), discovers mechanisms without supervision, and naturally extends to other kinetic systems where optimal control or excitation strategies are sought.

\section{Learning Kinetic Closures Through Differentiable Fluid Simulation}
\label{sec:closures}

The preceding example used differentiable simulation to discover what physical configurations produce novel behavior, that is, the optimization identified driver parameters that maximize a physics-based objective. The following example inverts this logic and asks the following question: Given kinetic behavior from Vlasov simulations, can we learn how to reproduce it within a computationally cheaper fluid framework? Both applications exploit the same programming infrastructure, but the first searches for interesting physics while the second learns to emulate it.

A central challenge in plasma modeling is constructing closure relations that capture kinetic effects when the Knudsen number is not small. In such regimes, fluid equations alone are insufficient because unresolved velocity-space dynamics introduce behavior that is nonlocal in both space and time. This section summarizes an approach in which differentiable programming enables learning a kinetic closure for nonlinear Landau damping by embedding a machine-learned dynamical variable directly into a fluid simulator, based on Joglekar and Thomas~\cite{joglekar_machine_2023}.

\subsection{The closure problem: nonlocality from resonant electrons}

Linear Landau damping, the collisionless damping of plasma waves through wave-particle resonance, can be incorporated into fluid models via a number of methods. While typically performed using a heat flux closure \cite{hammett_fluid_1990}, the simplest, and most useful for the application here, method is one that adds a wavenumber-dependent damping term to the momentum equation. However, when wave amplitudes become large, electrons become trapped in the wave potential, suppressing Landau damping through a nonlinear saturation mechanism. No closed-form expression exists for this nonlinear damping rate as a function of wave amplitude, collision frequency, and wavenumber. The situation becomes more complex for finite-length wavepackets such as those in the previous section where the back of the wave damps faster than the front due to the transit of resonant electrons.

This spatiotemporal nonlocality cannot be captured by any closure that depends only on local fluid quantities. The damping rate at a given location depends on the history of wave-particle interactions upstream which is ``hidden'' from the fluid description.

\subsection{Hidden-variable formulation}

To address this limitation, we introduce an auxiliary field $\delta(x,t)$ representing the population of resonant electrons. This hidden variable modifies the effective Landau damping rate and evolves according to its own transport equation. The coupled system, implemented in the differentiable fluid code ADEPT (Automatic-Differentiation-Enabled Plasma Transport), consists of the electron continuity and momentum equations, given by
\begin{align}
\partial_t n + \partial_x (u n) &= 0, \\
m n (\partial_t u + u \partial_x u) &= -\partial_x p + q n E + 2 m n \frac{\nu_L \ast u}{1 + \delta^2},
\end{align}
closed with an adiabatic equation of state. Here $\nu_L$ is the linear Landau damping kernel implemented as a spatial convolution, and the factor $(1+\delta^2)^{-1}$ provides sigmoidal suppression of damping when resonant electrons are present.

The hidden variable evolves according to
\begin{equation}
\partial_t \delta = v_{\mathrm{ph}} \, \partial_x \delta + \nu_g \frac{|E \ast \nu_L|}{1 + \delta^2},
\label{eq:delta_evolution}
\end{equation}
where the advection at $v_{\mathrm{ph}}$ reflects that resonant electrons travel at the phase velocity, and the source term is motivated by the wave-particle interaction term $E \partial_v f$ in the Vlasov equation. Landau showed that $\partial_v f \propto \nu_L$ for resonant particles, so the product $E \nu_L$ captures the local rate of resonant electron production. The saturation factor $(1+\delta^2)^{-1}$ ensures bounded growth, consistent with the finite energy available for particle trapping.

The growth rate coefficient $\nu_g$ depends on plasma parameters but has no known analytical form. It is therefore represented by a neural network,
\begin{equation}
\nu_g = \nu_g(k, \nu_{ee}, |\hat{E}_k|; \boldsymbol{\theta}),
\end{equation}
with output bounded to $10^{-3} < \nu_g < 10^3$ for numerical stability.

This construction encodes substantial physical knowledge. $\delta$ is transported at the correct velocity, grows in response to wave-particle interactions, and saturates at large amplitude. The neural network learns only the growth rate coefficient which is a single scalar function of three inputs, while all dynamics are governed by the coupled PDEs.

\subsection{Indirect supervision through differentiable simulation}

A key challenge is that $\delta$ has no direct counterpart in kinetic simulations. Rather, it is a reduced representation of velocity-space structure that cannot be extracted from Vlasov solutions. Training therefore requires \emph{indirect supervision} where the neural network parameters are optimized by comparing fluid and kinetic predictions of an observable quantity, with gradients backpropagated through the entire fluid simulation.

The loss function compares the time history of the first density Fourier mode and is given by
\begin{equation}
\mathcal{L} = \frac{1}{N_t} \sum_{t=250\omega_p^{-1}}^{450\omega_p^{-1}}
\left[ \log_{10}|\hat{n}_1^{\mathrm{fluid}}| - \log_{10}|\hat{n}_1^{\mathrm{Vlasov}}| \right]^2.
\end{equation}
The logarithmic comparison is essential because wave amplitudes can damp over 5--6 orders of magnitude. Without it, late-time behavior would be invisible to the optimizer.

Training data consists of Vlasov-Poisson-Fokker-Planck simulations spanning
\begin{align}
\nu_{ee} &\in [10^{-7}, 10^{-6}, \ldots, 10^{-3}], \\
k &\in [0.26, 0.28, \ldots, 0.40], \\
a_0 &\in [10^{-6}, 10^{-5}, \ldots, 10^{-2}],
\end{align}
where $\nu_{ee}$ is the collision frequency, $k$ the wavenumber, and $a_0$ the initial amplitude. From 2535 total kinetic simulations, 200 are subsampled for training (180) and validation (20). Each training simulation excites a single-wavelength wave in a periodic box of one wavelength.

The neural network is 3 layers deep and 8 nodes wide, with hyperbolic tangent activations, totaling only 160 parameters. This compact architecture likely suffices because the physics is encoded in the PDEs and the network learns only a correction factor.

\subsection{Results: from periodic boxes to wavepacket etching}

The trained model achieves a test loss of $10^{-2}$ on 2335 held-out simulations spanning the full parameter space, with training and validation losses of $5.8 \times 10^{-3}$ and $2.0 \times 10^{-2}$, respectively. The model correctly recovers linear Landau damping at small amplitudes ($a_0 = 10^{-6}$), where the hidden variable remains negligible at $\delta \approx 10^{-5}$, and reproduces nonlinear saturation at large amplitudes ($a_0 = 10^{-2}$), where $\delta$ grows and suppresses damping.

\begin{figure}[h]
    \begin{subfigure}{0.45\textwidth}
        \includegraphics[width=\textwidth]{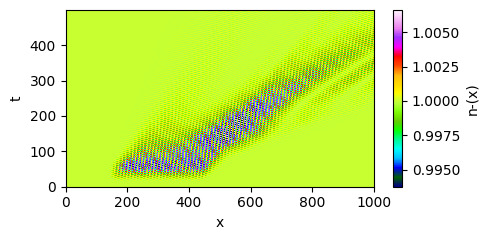}    
    \end{subfigure} 
    \\
    \begin{subfigure}{0.45\textwidth}
        \includegraphics[width=\textwidth]{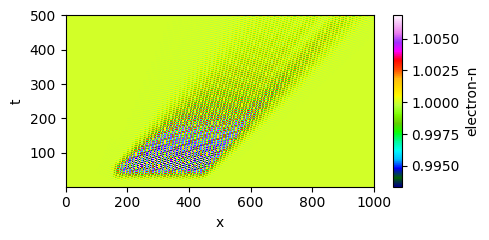}
    \end{subfigure}
    \\
    \begin{subfigure}{0.45\textwidth}
        \includegraphics[width=\textwidth]{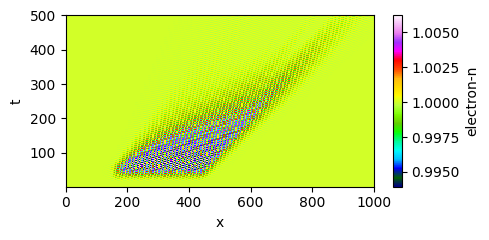}    
    \end{subfigure} 
    \\
    \begin{subfigure}{0.45\textwidth}
        \includegraphics[width=\textwidth]{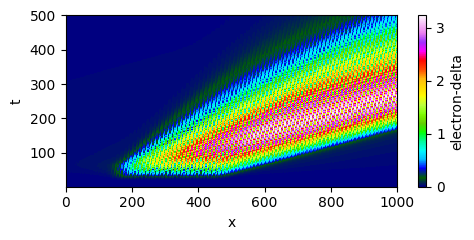}
    \end{subfigure}
    \\
\caption{Generalization to finite-length wavepackets in a domain 100$\times$ larger than the training geometry. (a) Vlasov simulation shows non-uniform damping with the rear eroding faster than the front. (b) Local closure produces uniform damping. (c) Learned hidden-variable closure reproduces the kinetic behavior without retraining. (d) The hidden variable $\delta$ grows at the wavepacket location and advects forward at the phase velocity, encoding spatial memory. From Joglekar and Thomas~\cite{joglekar_machine_2023}.}
\label{fig:wavepacket_etching}
\end{figure}

The critical test is generalization to geometries far outside the training distribution. Training used single-wavelength periodic systems; we evaluate the model on domains 100× larger containing finite-length wavepackets with open boundaries. In this geometry, kinetic simulations exhibit wavepacket etching—nonuniform damping where the rear of the packet erodes faster than the front due to resonant electrons streaming ahead of the wave. A local closure cannot capture this effect because damping at each location depends on the upstream history of wave-particle interactions. Figure \ref{fig:wavepacket_etching} illustrates this comparison: the local closure model produces spatially uniform damping, while the hidden-variable model reproduces the non-uniform erosion observed in kinetic simulations without retraining.

\subsection{Discussion}

This example demonstrates several principles central to differentiable programming for plasma physics:

\textbf{Learning what cannot be directly observed.} The hidden variable $\delta$ has no counterpart in kinetic theory; it is a phenomenological representation of velocity-space structure. Differentiable simulation enables training such variables through their effect on observables, a form of indirect supervision that would be impossible with offline supervised learning.

\textbf{Physics as inductive bias.} By embedding the neural network within a PDE system that encodes transport, saturation, and the correct characteristic velocity, the learned model generalizes far beyond its training distribution. The network learns a coefficient; the equations provide the structure.

\textbf{From idealized to realistic geometries.} Training on computationally cheap single-wavelength systems, then deploying to expensive multi-wavelength configurations, inverts the traditional workflow of running large simulations to generate training data for reduced models.

The hidden-variable approach extends naturally to other multiscale problems where kinetic effects must be represented in fluid models e.g. hot electron transport and nonlocal heat flow in inertial fusion plasmas, or kinetic corrections to MHD in the solar wind. In each case, the key is identifying a reduced representation that captures the essential nonlocality while remaining computationally tractable.

\section{Accelerating Experimental Diagnostics with Differentiable Programming}
\label{sec:thomson}

Many experimental plasma diagnostics rely on model-based inference where physical parameters are extracted by fitting a forward model to measured observables. These analyses are naturally inverse problems, yet in practice they are often solved using finite-difference gradients or manual iteration. This section summarizes an approach~\cite{milder_qualitative_2024} in which reverse-mode AD dramatically accelerates Thomson-scattering analysis while enabling qualitatively new measurement capabilities.

\subsection{Model-based diagnostics as inverse problems}

Let $\mathbf{D}$ denote experimentally measured data and $\mathcal{M}(\mathbf{p})$ a physics-based diagnostic model parameterized by plasma conditions $\mathbf{p}$. Parameter estimation seeks
\begin{equation}
\mathbf{p}^\ast = \underset{\mathbf{p}}{\mathrm{argmin}} \; \mathcal{L}(\mathbf{D}, \mathcal{M}(\mathbf{p})),
\end{equation}
where $\mathcal{L}$ quantifies the discrepancy between model and data. For Thomson scattering, $\mathbf{p}$ typically includes electron density $n_e$, electron temperature $T_e$, ion temperature $T_i$, ionization state $Z$, flow velocity, and instrumental normalization factors.

With finite differencing, each gradient evaluation costs $\mathcal{O}(N_p)$ forward model evaluations. For Thomson-scattering analysis with $N_p \sim 10$ parameters per spatial/temporal location and hundreds of locations per dataset, this cost becomes prohibitive. Traditional analyses therefore fit each location independently with manual intervention or restrict the parameter space.

\subsection{AD-enabled workflow transformation}

Reverse-mode AD computes the full gradient $\partial \mathcal{L}/\partial \mathbf{p}$ in time comparable to a single forward evaluation, regardless of $N_p$. This $\mathcal{O}(N_p) \to \mathcal{O}(1)$ scaling enables a fundamental workflow change. Data are split into batches of $J$ lineouts fit simultaneously by minimizing
\begin{equation}
\mathcal{L}_{\mathrm{batch}} = \sum_{j=1}^{J} \sum_{i=1}^{N_{\mathrm{pixel}}} 
\frac{\left[ D_{ij} - \mathcal{M}_i(\mathbf{p}_j) \right]^2}{D_{ij}},
\end{equation}
where each lineout has its own parameter vector $\mathbf{p}_j$. With finite differencing, batching is impractical because gradient cost grows as $J \times N_p$; with reverse-mode AD, gradient cost is nearly independent of this product.

The Thomson-scattering forward model computes the scattered power spectrum,
\begin{equation}
P_s(\lambda_s, \theta, \mathbf{p}) = A \, n_e \left( \frac{2\lambda_0}{\lambda_s^3} - \frac{1}{\lambda_s^2} \right) S(\lambda_s, \theta, \mathbf{p}),
\end{equation}
where $S$ is the spectral density function depending on electron and ion velocity distributions through susceptibility integrals. The full model includes convolution with instrumental response functions and integration over scattering angles. All components are implemented in JAX, rendering the complete model differentiable.

\subsection{Results: 140$\times$ acceleration with uncertainty quantification}

\begin{figure}[h]
\centering
\includegraphics[width=0.4\textwidth]{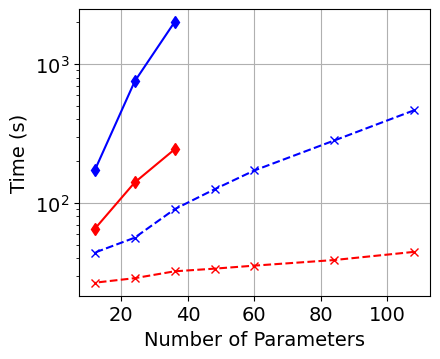}
\caption{Computation time versus number of fitting parameters (lineouts) for Thomson-scattering analysis using finite differencing (blue) and automatic differentiation (red), on CPU (solid) and GPU (dashed). Reverse-mode AD combined with GPU acceleration yields over two orders of magnitude speedup, with the advantage increasing for larger parameter counts.}
\label{fig:ts_timing}
\end{figure}

The combination of reverse-mode AD, batching, and GPU acceleration yields dramatic speedups, as shown in Fig.~\ref{fig:ts_timing}. Analysis of a temporally resolved dataset that previously required 90 minutes for 20 lineouts using finite differencing on CPU now completes in 11 minutes for 360 lineouts which is a $>$140$\times$ acceleration per lineout. This speedup arises from reverse-mode AD ($\sim$10$\times$), GPU parallelization ($\sim$10$\times$), and batching efficiency ($\sim$1.4$\times$), with no approximations to the physics model.

This acceleration has immediate practical consequences. Every pixel can be analyzed rather than sparse sampling, near real-time feedback becomes possible for adaptive experiments, and complete datasets can be processed rapidly without manual intervention.

\begin{figure*}[ht!]
\centering
\includegraphics[width=0.8\textwidth]{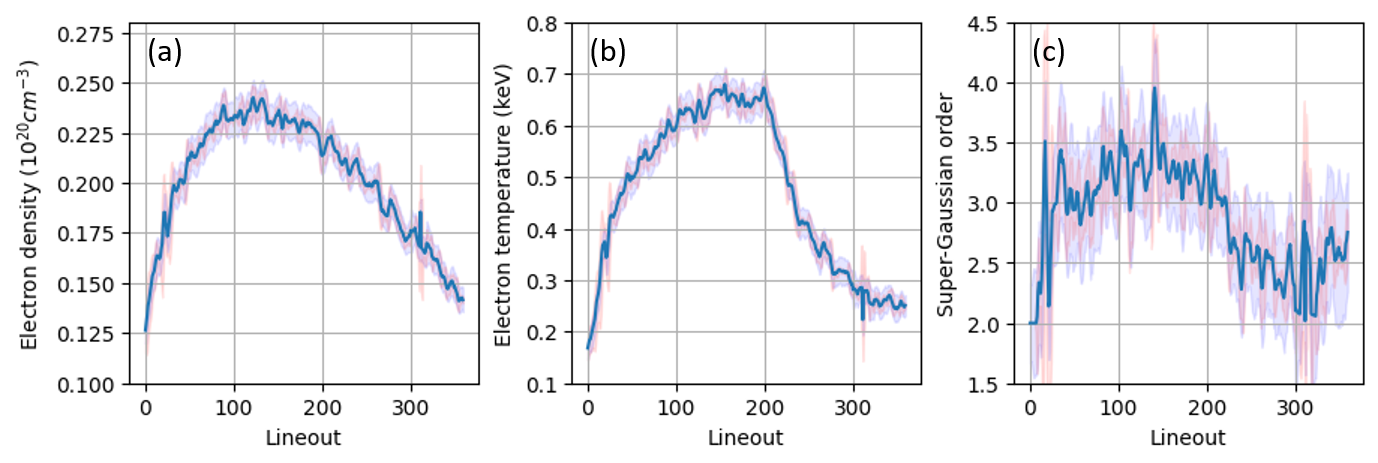}
\caption{Plasma conditions extracted from temporally resolved Thomson-scattering data: (a) electron density, (b) electron temperature, and (c) super-Gaussian order of the electron distribution. Blue curves show fitted values at every pixel; shaded regions indicate $3\sigma$ uncertainty from the Hessian (blue) and from the standard deviation within a 5-pixel moving window matching the diagnostic resolution (red). The two uncertainty estimates are in good agreement. From Milder \textit{et al.}~\cite{milder_qualitative_2024}.}
\label{fig:ts_conditions}
\end{figure*}

AD also enables efficient uncertainty quantification through the Hessian $\mathbf{H} = \nabla \nabla \mathcal{L}$. Under locally Gaussian assumptions, the covariance matrix is $\boldsymbol{\Sigma} = 2\mathbf{H}^{-1}$. With finite differencing, Hessian computation scales as $\mathcal{O}(N_p^2)$; reverse-mode AD reduces this to $\mathcal{O}(N_p)$ or $\mathcal{O}(1)$ using forward-over-reverse composition. Figure~\ref{fig:ts_conditions} shows plasma conditions extracted at every pixel with uncertainty bounds computed via this approach, validated against moving-window statistics.

\subsection{High-dimensional inference: measuring distribution functions}

The most striking application is fitting quantities previously computationally inaccessible. In angularly resolved Thomson scattering~\cite{milder_qualitative_2024}, the spectrum is measured versus both wavelength and scattering angle, enabling direct reconstruction of the electron velocity distribution $f_e(v)$.

Rather than assuming a Maxwellian, $f_e$ can be represented as $N_v$ discrete values at specified velocities, all treated as free parameters. With $N_v = 256$ plus plasma conditions and instrumental factors, optimization is intractable with finite differencing but accessible with reverse-mode AD because the full gradient is obtained in a single backward pass. Analysis that previously required 8 hours with 64 velocity points now completes in 15 minutes with 256 points giving a 30$\times$ speedup with 4$\times$ improved resolution.

This transforms Thomson scattering from measuring bulk parameters into directly probing kinetic physics through distribution function shapes. Non-Maxwellian features such as accelerated tails, depleted populations, multi-component structure may become measurable.

\subsection{Discussion}

The Thomson-scattering application illustrates that AD enables qualitatively new measurements, not merely accelerating existing analyses. Any diagnostic whose forward model can be implemented in a differentiable framework benefits from the same acceleration and expanded parameter space e.g. proton radiography, x-ray spectroscopy, and interferometry.

\section{Inverse Design of Spatiotemporal Laser Pulses}\label{sec:design}

The preceding sections applied differentiable programming to discover novel physics, learn kinetic closures, and accelerate experimental inference. A fourth application, inverse design, completes this picture by using gradient-based optimization to determine what input structure produces a desired output. This section summarizes work by Miller \textit{et al.}~\cite{miller_spatiotemporal_2025} demonstrating that automatic differentiation enables the design of structured laser pulses with spatiotemporal features advantageous for nonlinear optics and plasma physics.

\subsection{Inverse design as an optimization problem}

Laser pulses with coupled space-time structure offer additional degrees of freedom for optimizing laser-based applications. The ``flying focus,'' for example, produces a peak intensity that moves independently of the group velocity over many Rayleigh ranges~\cite{froula_spatiotemporal_2018}. While such known structures have proven useful for plasma and nonlinear optical processes, the optimal structure for a particular application may combine existing approaches or take an entirely new form.

Traditional forward design evaluates how a known structure affects an outcome. Inverse design reverses this logic and asks, given a desired outcome, what structure achieves it? This formulation maps naturally onto gradient-based optimization. Let $\mathbf{p}$ denote the parameters specifying a laser pulse in the near field, and let $\mathcal{S}(\mathbf{p})$ represent the far-field evolution computed by a physics-based propagation model. A loss function $\mathcal{L}$ quantifies the discrepancy between the simulated outcome and the target. The inverse problem seeks
\begin{equation}
    \mathbf{p}^* = \underset{\mathbf{p}}{\mathrm{argmin}} \; \mathcal{L}\bigl(\mathcal{S}(\mathbf{p}), \mathcal{T}\bigr),
\end{equation}
where $\mathcal{T}$ represents the target behavior. Reverse-mode AD provides exact gradients $\partial \mathcal{L}/\partial \mathbf{p}$ at cost independent of the number of parameters, enabling optimization over high-dimensional pulse representations.

\begin{figure*}[!t]
    \centering
    \includegraphics[width=\textwidth]{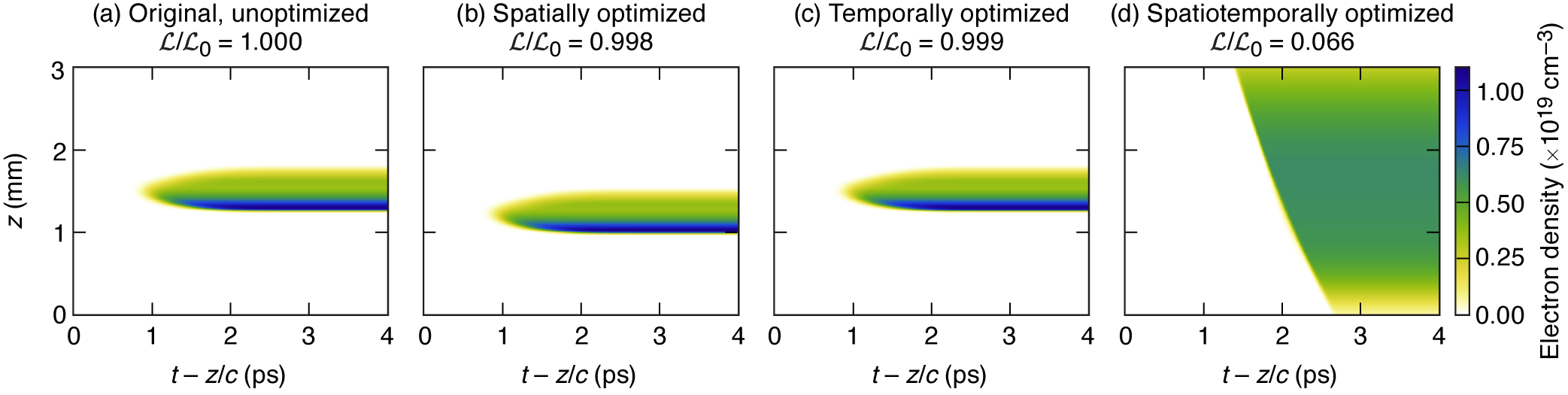}
    \caption{Inverse design of a laser pulse to generate a uniform plasma column. Electron density as a function of $z$ and $t-z/c$ for (a) an unoptimized pulse focused to the nominal focal plane, (b) optimization of spatial structure only, (c) optimization of temporal structure only, and (d) full spatiotemporal optimization. The ratio of final to initial loss $\mathcal{L}/\mathcal{L}_0$ is indicated for each case. Only coupled space-time control achieves significant improvement, reducing the loss by 93\%. Adapted from Miller \textit{et al.}~\cite{miller_spatiotemporal_2025}.}
    \label{fig:spatiotemporal}
\end{figure*}
\subsection{Differentiable pulse propagation}

The far-field evolution is modeled using the unidirectional pulse propagation equation (UPPE), a reduced form of Maxwell's equations that neglects backward-propagating waves while retaining dispersion to all orders and nonlinear effects including ionization~\cite{kolesik_unidirectional_2002}. For a cylindrically symmetric pulse propagating in the $z$-direction, the UPPE in spectral space takes the form given by
\begin{equation}
    \frac{\partial \tilde{E}}{\partial z} - i K(k_r, \omega) \tilde{E} = \tilde{S}(k_r, \omega, \tilde{E}),
\end{equation}
where $K = k_z - \omega/v_f$ accounts for diffraction and dispersion in a frame moving at velocity $v_f$, and $\tilde{S}$ contains nonlinear source terms including the Kerr effect, plasma current, and ionization energy loss.

The solver is implemented in JAX, rendering the complete propagation from near field to far field differentiable with respect to all pulse parameters. The near-field amplitude is parameterized through radial functions for intensity, central frequency, and pulse duration, while the phase is expanded in products of frequency powers and Zernike polynomials. This parameterization is sufficiently general to represent diverse spatiotemporal structures while remaining tractable for optimization.

\subsection{Application: three structured pulse designs}

The framework is applied to three design problems of increasing complexity, each achieving at least a factor of 15 reduction in loss.

\subsubsection{Longitudinally uniform intensity}

Axicons and axiparabolas produce extended focal regions but typically exhibit longitudinal intensity modulations. The first optimization seeks a monochromatic beam that maintains constant on-axis intensity over a 1.7~cm region. Only spatial structure (amplitude and monochromatic phase) is optimized. The converged solution achieves a nearly flat intensity profile, outperforming a standard axiparabola by a factor of 7 in uniformity. The learned phase coefficients closely resemble those of an axiparabola but include additional spherical aberration terms that prove essential for uniformity.

\subsubsection{Spatiotemporally uniform flying focus}

Many applications require an intensity peak that travels at a prescribed velocity with constant duration, spot size, and amplitude over an extended region. The second optimization targets a superluminal ($1.005c$) intensity peak with 40~fs duration and 10~$\mu$m spot size maintained over 8~mm (27 Rayleigh ranges). Full spatiotemporal optimization of amplitude and phase is employed.

The converged pulse exhibits complex near-field structure unlikely to emerge from forward design. The optimization results in a radially varying central frequency, duration, and amplitude combined with chromatic focusing and spherical aberration. The optimized spot size varies by only 6\% over the target region, compared to a factor of 2.4 variation for a conventional axiparabola-echelon combination achieving the same focal velocity.

\subsubsection{Uniform plasma column generation}

The third optimization incorporates the full nonlinear physics of ionization. The target is a 1.6~mm long, 4~$\mu$m radius column of uniform plasma density ($6 \times 10^{18}$~cm$^{-3}$) in hydrogen gas. Because ionization causes refraction and dispersion, the optimal pulse for creating uniform plasma differs from one producing uniform intensity in vacuum.

The optimization discovers a flying-focus configuration in which the central frequency increases quadratically with radius, producing a superluminal ($1.14c$) intensity peak. Critically, neither purely spatial nor purely temporal optimization achieves significant improvement. Spatial-only optimization reduces the loss by 0.2\%, temporal-only by 0.1\%, while full spatiotemporal optimization reduces it by 93\%. This result demonstrates that coupled space-time structure provides capabilities inaccessible to either spatial or temporal shaping alone.

\subsection{Discussion}

This application illustrates several principles that complement the preceding sections. First, inverse design and discovery (Section~III) are conceptually symmetric. Discovery asks what physics emerges from optimized inputs, while inverse design asks what inputs produce desired physics. Both exploit the same AD infrastructure.

Second, the plasma column example demonstrates that nonlinear physics can qualitatively change optimal designs. A pulse optimized for uniform intensity in vacuum would not produce uniform plasma; the optimization must account for the coupled evolution of the electromagnetic field and the ionizing medium.

Third, the utility of differentiable simulation depends on whether optimized structures can be realized experimentally. The complex near-field profiles emerging from optimization motivate ongoing development of spatiotemporal pulse shaping techniques including metasurfaces and programmable optics. Conversely, constraints reflecting experimental capabilities could be incorporated directly into the loss function to ensure realizability.

The framework extends naturally to other inverse design problems in laser-plasma physics, e.g. optimizing pulse shapes for laser wakefield acceleration, designing drivers for terahertz generation, or tailoring ionization profiles for plasma waveguides.

\section{Conclusion}\label{sec:conclusion}
The examples presented in this article demonstrate that differentiable programming provides a practical framework for reformulating plasma physics problems as inverse problems. Rather than prescribing parameters and interpreting outcomes, one poses questions directly in terms of objectives. What plasma state best explains a diagnostic measurement? What closure reproduces kinetic behavior at fluid scales? What driver configuration maximizes a nonlinear response?

Four insights emerge from these applications. First, reverse-mode AD makes high-dimensional optimization tractable. Fitting velocity distributions with over 1000 parameters, or learning continuous functional relationships through neural networks with hundreds of weights, becomes routine rather than prohibitive. Second, this approach does not require abandoning first-principles physics. The neural network in the closure example learns only a single coefficient while all dynamics remain governed by physically motivated transport equations. Such models can be viewed as physics-based simulations augmented with learnable components rather than black-box replacements. Third, the same computational infrastructure serves theory, simulation, and experiment. The AD tools that accelerate diagnostic analysis also enable discovery of nonlinear wave interactions and learning of multiscale closures. Fourth, inverse design and physics discovery are conceptually similar in that both use optimization to navigate parameter spaces that would be intractable to explore manually, demonstrating that the same computational infrastructure serves both theoretical exploration and experimental planning.

Notably, the kinetic discovery application identified a previously unknown superadditive regime in which interacting wavepackets sustain electrostatic energy longer than either would in isolation. This was a genuine physics discovery enabled by gradient-based optimization of a differentiable simulation.

Important limitations remain. Extracting interpretable physical laws from trained neural networks is challenging, though symbolic regression applied to universal approximators offers a partial solution. Memory requirements for reverse-mode AD can be substantial for long simulations, necessitating checkpointing strategies. Finally, while AD provides exact gradients of discretized equations, care must be taken that these reflect the underlying continuous physics~\cite{metz_gradients_2022}.

Looking forward, the closure learning approach extends naturally to effects relevant to fusion and astrophysics applications. The diagnostic framework applies to any forward model implementable in a differentiable language.  Combining differentiable simulation with symbolic regression offers a path toward automated hypothesis generation, effectively using optimization to identify candidate mechanisms for subsequent theoretical analysis. As differentiable programming tools mature, we anticipate these methods will complement the theoretical insight and experimental intuition that remain essential to plasma physics. 

Indeed, differentiable programming for plasma physics is emerging as a broader research direction beyond the applications presented here. Carvalho et al. [ref, ref] recently demonstrated that differentiable kinetic simulators can infer collision operators directly from particle-in-cell data, recovering advection and diffusion coefficients without assuming a functional form a priori — an approach that extends naturally to regimes where no closed-form collision theory exists. In magnetic confinement fusion, the stellarator optimization community has been an early and prolific adopter of AD: the FOCUSADD code~\cite{mcgreivy_optimized_2021} and the DESC optimization suite~\cite{panici_desc_2022,conlin_desc_2022,dudt_desc_2022} use reverse-mode AD to optimize coil geometries and MHD equilibria over hundreds to thousands of parameters, a setting where hand-derived adjoints are impractical to maintain as objectives evolve. That differentiable programming is independently being adopted for operator discovery in kinetic theory, high-dimensional design in magnetic confinement, and the diagnostic and laser-plasma applications presented here suggests that the approach addresses a genuine and widespread need across plasma physics.

\section*{Acknowledgments}

This material is based upon work supported by IFE COLoR under U.S. Department of Energy Grant No. DE-SC0024863, Office of Fusion Energy Sciences under Award Numbers DE-SC0021057, US DOE National Nuclear Security Administration (NNSA) Center of Excellence under Cooperative Agreement No. DE-NA0003869, the Air Force Office of Scientific Research under Award Number FA-9550-22-1-0400, and by the Department of Energy National Nuclear Security Administration under Award Number DE-NA0004144, the University of Rochester, and the New York State Energy Research and Development Authority. This research used resources of the National Energy Research Scientific Computing Center, a DOE Office of Science User Facility supported by the Office of Science of the U.S. Department of Energy under Contract No. DE-AC02-05CH11231 using NERSC award FES-ERCAP0026741.

\section*{Author Contributions}
A. J. conceived the overall project and conceived the discovery and closure applications. A. J. implemented the kinetic and fluid solvers. A. J. and A. T. performed the data analysis. A. M. and A. J. implemented the differentiable Thomson-scattering model and performed the data analysis. K. M. and J. P. conceived the spatiotemporal pulse design application. K. M. implemented the differentiable pulse propagation model and performed the inverse design optimizations. A. T. and D. F. acquired funding. A. T., D. F., and J. P. supervised the project. All authors contributed to writing and editing the manuscript.

\section*{Data Availability}
Data sharing is not applicable to this article as no new data were created or analyzed in this study.

\bibliographystyle{apsrev4-2}
\bibliography{pop}

\end{document}